\documentclass[]{aa}  
\usepackage{graphicx}
\usepackage{txfonts}
 \usepackage{amsmath}
\usepackage[]{hyperref}
\usepackage[modulo, switch]{lineno}

\begin{document} 

   \title{Estimating solar radiation environment extremes}

   \author{Konstantin Herbst
          \inst{1, 2}
          \and
          Athanasios Papaioannou\inst{3}
          }

   \institute{Centre for Planetary Habitability (PHAB). Faculty of Mathematics and Natural Sciences. Department of Geosciences. University of Oslo. Sem Sælands vei 2A Blindern. Norway\\
        \email{konstantin.herbst@geo.uio.no}
        \and
        Institut f\"ur Planetenforschung (PF). Deutsches Zentrum f\"ur Luft- und Raumfahrt (DLR). Rutherfordstr. 2, 12489 Berlin. Germany
        \and
             Institute for Astronomy. Astrophysics. Space Applications and Remote Sensing (IAASARS). National Observatory of Athens. I. Metaxa \& Vas. Pavlou St. 15236 Penteli. Greece\\
             \email{atpapaio@astro.noa.gr}
             }

   \date{}
 
  \abstract
   {}
   {Extreme Solar Energetic Particle Events (ESPEs) were identified almost a decade ago, providing context for super events unleashed by our host star, the Sun. Their assumed solar origin drives the question of their ``worst-case" impact, which could be profound, multifaceted, and devastating for our technological society.}
   {A methodology that directly relates the soft X-ray flux $F_{SXR}$ of the driving solar flare of a Solar Energetic Particle event to its ``worst-case" integral fluence spectrum has recently been proposed by \cite{2023A&A...671A..66P}. We employ this method to the ESPEs that have been confirmed in cosmogenic radionuclide records up to date, retrieve their ``worst-case" integral spectrum, and compare the latter to the actual -- independently obtained -- recent reconstructions based on the radionuclide records.}
   {We first show that our method allows us to estimate the integral fluence spectra of one of the paleo events, i.e., AD774/775, one of the strongest ESPEs found within the cosmogenic radionuclide records so far. We then implement a mean ESPE utilizing four confirmed paleo ESPEs (i.e., AD993/994, AD774/775, 660 BCE, and 7176 BCE) and test the resulting spectrum against the estimated one. Finally, we test the same methodology for a series of strong SEPs recorded on the Earth's surface as Ground Level Enhancements (GLEs). In all investigated cases, the recent re-calibration of $F_{SXR}$ by \cite{hudson2024} is considered.}
   {We conclude that the methodology can adequately estimate the ``worst-case" integral fluence spectra for both strong and extreme SEP events, quantifying their impact up to an integral energy of $\sim$ E $>$ 1 GeV. }


   \maketitle
%
\section{Introduction} \label{sec:intro}
Solar Energetic Particle (SEP) events represent one of the most dynamic and impactful phenomena in space physics \citep{2023FrASS..1054266R}. These events occur when particles, primarily protons, electrons, and heavier ions, are accelerated to near-relativistic speeds by solar processes \citep{1999SSRv...90..413R}. As a result, the energies of SEP events span from a few keV to a few GeV \citep[see discussion in][]{2023A&A...671A..66P}. The relativistic high-energy tail of SEPs is occupied by Ground Level Enhancements (GLEs); in these events, particles are energetic enough \citep[E$\ge$300MeV; see details in][]{2021SoPh..296..129M}, allowing them to penetrate Earth's magnetic field and reach the lower atmosphere, producing secondary particles detectable on the ground, usually by neutron monitors  \citep[see a recent review in][]{macau_mods_00083703}. Two primary mechanisms drive SEP events: solar flares and coronal mass ejections (CMEs) \citep{klein2017acceleration}. On the one hand, solar flares, which are intense bursts of radiation emanating from releasing magnetic energy in the Sun’s atmosphere, can accelerate particles through magnetic reconnection processes \citep{2002SSRv..101....1A}. On the other hand, CMEs involve large expulsions of plasma and magnetic field from the Sun's corona outward into the interplanetary space \citep{2023FrASS..1064226H}. The interaction of these masses expelled from the Sun with the underlying ever-present solar wind creates shock waves that can also accelerate particles to high energies as they propagate through the heliosphere via, e.g., diffusive shock acceleration (DSA) \citep{desai2016large}. However, solar flares and CMEs do not evolve in isolation but rather in concert. In particular, solar flares as sudden, intense bursts of electromagnetic energy have a prominent soft X-ray (SXR) emission. The intensity of the SXR emission often correlates with the flare's strength. CMEs, which involve the ejection of massive amounts of solar plasma and magnetic fields into space, frequently accompany solar flares. While not all flares lead to CMEs, the most powerful flares are typically followed by significant CMEs \citep{2006ApJ...650L.143Y}. Both flares and CMEs are governed by the same magnetic processes in the solar atmosphere \citep{2001ApJ...559..452Z,2021LRSP...18....4T}, and SXR emission often serves as an early indicator of the flare's intensity and the likelihood of a CME \citep[see][and references therein]{2016ApJ...833L...8T, 2024A&A...690A..60P}. Details on the SXR-CME relation are added in Appendix \ref{appendix:A}. As a result, most SEP events are associated with both of these drivers \citep[see, e.g.][]{2010JGRA..115.8101C,2016JSWSC...6A..42P}. 

Although modern GLE events \citep{2012SSRv..171...61N,2023SpWea..2103334W} are often referred to as extreme solar events, a handful of sudden, much more extreme increases have been identified in the cosmogenic radionuclide records of $^{14}$C, $^{10}$Be, and $^{36}$Cl \citep[e.g.,][]{2012Natur.486..240M, mekhaldi2015multiradionuclide}. Because of the strength of these increases, several possible explanations, including gamma-ray bursts \citep{10.1093/mnras/sts378} and cometary impact \citep{LiuEA2014}, have been discussed in the literature. Nowadays, however, we know that these spike-like increases were caused by impacting solar energetic particles \citep{usoskin2013ad775, https://doi.org/10.1002/grl.50222, CliverEA2014} that form the actual extreme solar particle events \citep[ESPEs, see the detail review by][]{2023SSRv..219...73U}. By nature, these extreme events have not been directly observed/measured. Consequently, their energy spectrum is at times represented by a scaled spectrum of the strongest ever observed hard-spectrum SEP event: GLE05 that occurred on 23 February 1956 \citep[see details in][]{2023A&A...671A..66P}.

\begin{figure}[!t]
\vspace{-0.2cm}
\includegraphics[width=\columnwidth]{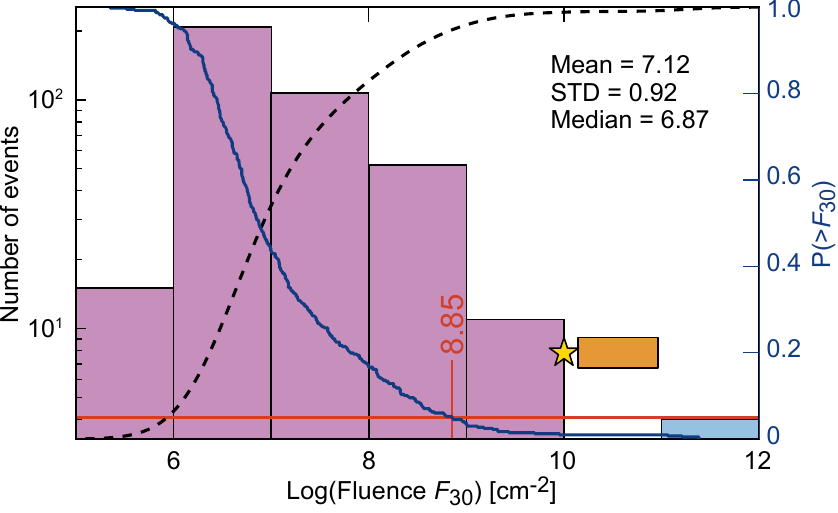}
\caption{Statistical distribution of the F$_{30}$ proton fluence obtained using a total of 392 SEP events recorded from 1955 to 2023 onboard spacecraft (magenta bars) and the ESPEs of the past (light blue bar). The black dashed line corresponds to the cumulative distribution function (CDF). The blue trace (right ordinate axis) corresponds to the empirical Complementary CDF (CCDF) and gives the probability P($>F_{30}$) of having a SEP event with a log‐fluence above a certain value F \citep[see e.g.][]{Ross2014}. The horizontal
red line indicates the 5\% chance of exceeding the value indicated in red. The plot imprints the mean, median, and standard deviation of the entire sample for the log-fluence values. The y-axis is displayed as a log scale so that the whole range of log-fluence values can be displayed without the small ESPE values being compressed. The star symbol corresponds to the Carrington log-fluence value of 10$^{10}$cm$^{-2}$ \citep{2013JSWSC...3A..31C}. The orange rectangle depicts the lack of log-fluence values from 10$^{10-11}$cm$^{-2}$.}
\label{fig:frequency}
\end{figure}

One of the first and -- so far -- strongest events discovered in the cosmogenic radionuclide records, the AD774/775 event, was identified in 2012 \citep{2012Natur.486..240M} and immediately got a lot of attention, opening a new pathway for investigating physical processes on the Sun \citep{2021GeoRL..4894848U}. In the 12 years since that milestone starting point, three more ESPEs have
been confirmed in all three cosmogenic radionuclide records \citep[see details in][]{2022NatCo..13..214P,2023JGRA..12831186K}: the 993/994 CE \citep{2013NatCo...4.1748M}, the 660 BCE \citep{2019PNAS..116.5961O}, and the 7176 BCE \citep{2022NatCo..13.1196B} events. At present, these four events constitute the high energy/low probability tail in the SEP distribution (see, e.g., Fig.\ref{fig:frequency}). With a fluence of protons at E>30 MeV (F$_{30}$) estimated to be 1.5-2 orders of magnitude larger than the largest GLE on record (GLE05), the AD774/775 event is the largest SEP event described in detail so far. ESPEs are significantly larger than any SEP and/or GLE events detected since the 1950s \citep{2023JGRA..12831186K}, implying that such events pose an underestimated threat to our society.
F$_{30}$ can be determined from spacecraft data obtained between 1955 and 2023. In particular, a total of 392 SEP events reaching such an integral energy have been listed until today \citep[see details of this catalog in][]{papaioannou2024}. The histogram in Fig.~\ref{fig:frequency} displays the distribution of the logarithm of the F$_{30}$ values for all of these events. As can be seen, no event in the instrumental era (magenta bars) shows an F$_{30}$ value that exceeds the estimated Carrington event value (10$^{10}$ cm$^{-2}$)\footnote{see \cite{2013JSWSC...3A..31C}} (star symbol). The radionuclide-based F$_{30}$ estimates are shown as the last bar (in cyan), representing the four confirmed events (7176 BC, 660 BC, 774 AD, and 993 AD). The red horizontal line indicates the 5\% chance of exceeding the value imprinted in red, indicating the tail of the distribution. It shows that for only 5 out of 100 times, a log-fluence F$_{30}$ larger than 8.85 (cm$^{-2}$) may occur at our Sun. On the other hand, this means that 95 out of 100 times, the log-fluence will be below this value. This is further corroborated by the dotted black line, providing the CDF of the log-fluence. It indicates that the probability (P(>F$_{30}$)) that the log-fluence F$_{30}\leq$ 8.85 (cm$^{-2}$) is 0.95 (right hand ordinate) \citep[see also the relative discussion in][]{2022JSWSC..12...24P}. The observational gap between instrumental and paleo SEPs (orange rectangle in Fig.\ref{fig:frequency}) means we cannot conclude whether these two sets of observations result from the same underlying distribution and, therefore, are formed by the same physical processes.

Based on the existing $^{14}$C, $^{10}$Be, and $^{36}$Cl data, it can be hypothesized that no events larger than AD774/775 will be found in the radionuclide records from the last ten millennia \citep[e.g., ][]{MiyakeEA2020b}. However, most observed GLEs show `soft' spectra, with higher fluences of lower energy particles. Thus, previously unidentified extreme soft-spectrum events might hide in the cosmogenic radionuclide records (i.e., $^{36}$Cl).

\section{From soft X-rays to ``worst case" scenario particle fluences}
Recently, \cite{2023JGRA..12831186K} presented new, calibrated, multiproxy reconstructions of the integral fluence spectrum of all known ESPEs. In turn, this work provides precise numerically derived ready-to-use values of the reconstructed ESPE fluences $F(>E)$ (see their Table 3). At the same time, the tight connection between solar eruptive events and SEPs has been explored and discussed in numerous works \citep[e.g.][]{2010JGRA..115.8101C,2016JSWSC...6A..42P, desai2016large}. 

Most recently, a method that allows inferring the ``worst case" fluence spectrum of a SEP event solely utilizing the magnitude of the associated solar flare in terms of Soft-X-rays (SXRs) has been proposed by \cite{2023A&A...671A..66P}. The ``worst case" fluence is derived from the upper limit estimation of each fluence value at each integral energy of interest, following the methodology detailed in \cite{2023A&A...671A..66P}. The latter is based on \textit{upper limit} scaling relations that convert the SXR flux to peak proton flux ($I_{P}(>E)$) and fluence ($F_{P}(>E)$) for a set of integral energies from E$>$10 MeV to E$>$100 MeV. In particular, these authors utilized a carefully selected sample of 65 SEP events that extended from >10 to >100 MeV and identified scaling laws of SEPs to SXRs for each integral energy (i.e., >10, >30, >60, and >100 MeV). In addition, they converted the peak proton fluxes to fluences for each of the integral energies. An inverse power-law fit is assumed to fit the resulting SEP fluences from >10 to >100 MeV. Once established, this is extended to higher integral energies (i.e., E$>$200 MeV and E$>$430 MeV). \citet{2023A&A...671A..66P} found that $F_{P}(>E)$ scales with the SXR flux as $F_{SXRs}^{5/6}$. Thus, the integral fluence spectrum can be estimated directly when the $F_{SXR}$ is known.

Table 3 of \cite{2023JGRA..12831186K} provides the integral fluences $F_{P}(>30)$, $F_{P}(>60)$, $F_{P}(>100)$, $F_{P}(>200)$, $F_{P}(>300)$, $F_{P}(>600)$, and $F_{P}(>1000)$ for the ESPEs recorded around AD993/994, AD774/775, 660 BCE, and 7176 BCE. Thus, the combination of these integral fluences for each integral energy leads to a well-specified range of values per energy bin. In turn, these estimates can be used to construct the upper and lower limits of the integral fluence as a function of integral energy as deduced by measurements (proxies) alone.
\begin{figure}[!t]
\centering
\includegraphics[width=\columnwidth]{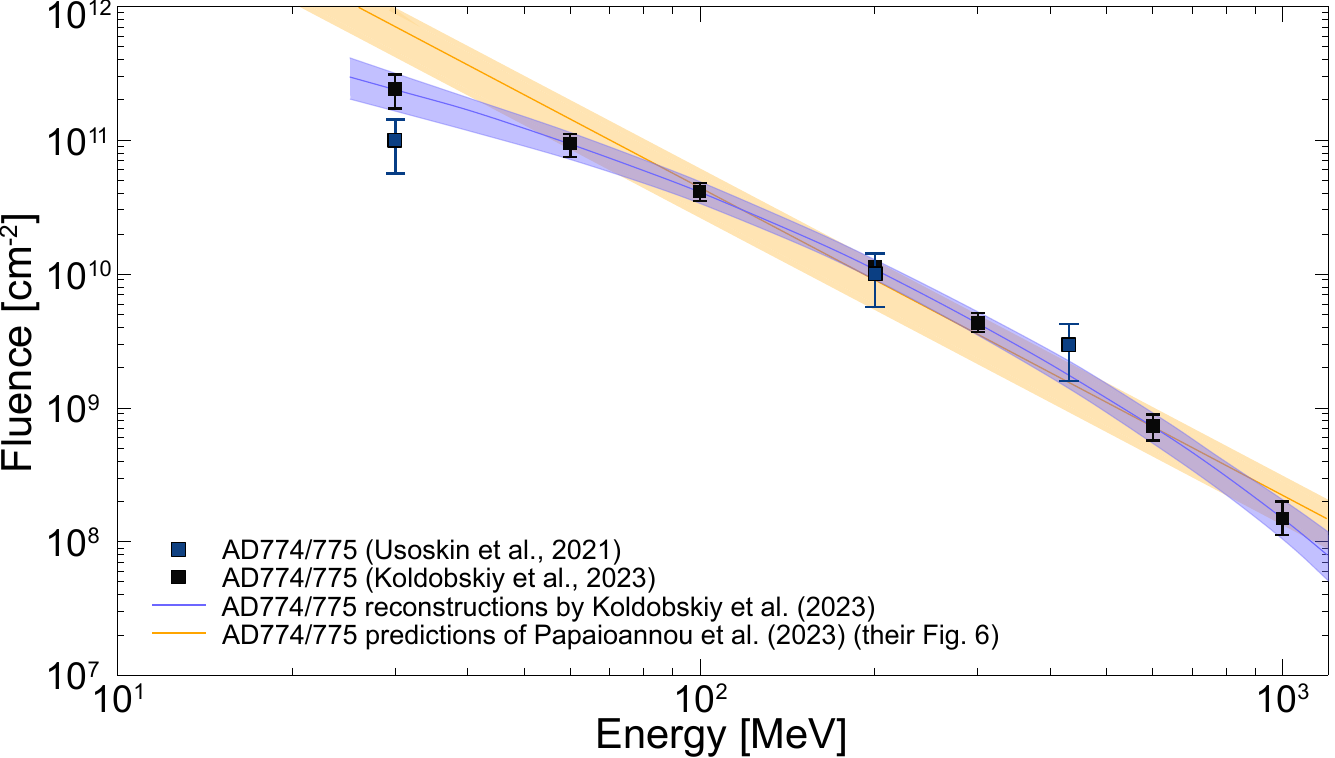}
\caption{The integral fluence spectra of the AD774/775 event.  Blue and black filled squares display the estimations based on the cosmogenic radionuclide data \citep[][, respectively]{usoskin2021strongest,2023JGRA..12831186K}. While the magenta ribbon reflects the spectrum derived by \cite{2023JGRA..12831186K}, the orange ribbon is obtained as a ``worst case" upper limit integral fluence spectra based on an X600 solar flare by employing the method outlined in \cite{2023A&A...671A..66P}.}
\label{fig:1}
\end{figure}

In the case of the AD774/775 event, \cite{2022LRSP...19....2C} identified the associated SXR flux as X400$\pm$200 \citep[see also][]{2023A&A...671A..66P,hudson2024}. This SXR flux was identified as follows: \citet{2020ApJ...903...41C} (i.e., their Fig.~7) began with a reduced-major-axis (RMA) fit to a scatter plot of modeled >200 MeV fluences taken from \citep{2018JSWSC...8A...4R} for hard-spectrum GLEs vs. the peak intensities of their associated SXR flares. They then added two points for the 1956 GLE based on the estimated range of the peak SXR intensity (X10-X30) of its associated flare (as inferred from white-light and radio observations) and its >200 MeV fluence \citep{usoskin2020revised}. Through these points, they extrapolated lines parallel to the RMA fit to the modeled >200 MeV fluence for the AD774 SEP event to obtain an estimate of X285$\pm$140 for the AD774 flare. Re-scaling the SXRs according to \citep{hudson2024} shifts the SXR of AD774 from X285$\pm$140 to X400$\pm$200. It follows that the worst-case upper limit SXR flux is X600. With the help of the method described in \citet{2023A&A...671A..66P}, therefore, we obtained the ``worst case" integral fluence spectra utilizing the X600 $F_{SXR}$ (orange solid line embedded in an envelope depicting the 1-$\sigma$ error; parameters of the fit are available in Table 3 of \cite{2023A&A...671A..66P}). As shown in Fig. \ref{fig:1}, the so-derived spectrum and the independently obtained range of integral fluence values from \cite{2023JGRA..12831186K} and \citet{usoskin2021strongest} can be directly compared. As can be seen, our AD 774/775 estimate aligns well with the fluence values derived by \citet{usoskin2021strongest} (blue symbols) and \citet{2023JGRA..12831186K} (black symbols). The underlying assumption of the aforementioned studies is that flares (denoted by their SXR measurements) drive SEPs. However, the lack of a universal correlation between spectral hardness and SXRs under a flare-only acceleration model \citep[see e.g.][]{2015SoPh..290..819T} suggests the need for an additional mechanism, such as CME-driven shocks -- supporting a dual-mechanism model \citep{2016JSWSC...6A..42P, desai2016large}. Considering the differences, the obtained alignment is promising. The spectral shape derived by \citet{2023JGRA..12831186K} is highlighted in purple. Given the fact that the AD774/775 event is the strongest ESPE event identified to this day, it represents an upper-limit "worst-case" event, demonstrating what our host star the Sun is capable of producing \citep{2023A&A...671A..66P}. The study of \citet{2023JGRA..12831186K} provides the scaled spectrum of the AD774/775 event with a 1-$\sigma$ spread. This is incorporated in the direct comparison of Fig.~\ref{fig:1}. Although there is an apparent roll-over energy for ESPEs (here at E$>$300 MeV), our simple inverse power-law representation (within the errors) gives a quite practical representation between E$>$60 MeV and E$>$1 GeV.
Nonetheless, the spectral shape employed (i.e., inverse power-law) is restrictive for both the softening of the higher energies and the underestimation of the lower energies. As a result, while a realistic spectrum for SEPs is expected to be a double power or have an exponential roll-over (see details in Appendix \ref{appendix:B}), here we consider a single power-law (PL) approximation in the energy range of E>10 MeV to E>100 MeV, extrapolated to E>1 GeV \citep[see also the relevant discussion in][]{2023A&A...671A..66P}. Nonetheless, in the work at hand, it was made evident that the simple PL can provide a reasonable estimation of the fluence spectra from E>60 MeV to E>1 GeV, which in turn provides a quick and practical representation of the ``worst-case" conditions.
\begin{figure}
\centering
\includegraphics[width=\columnwidth]{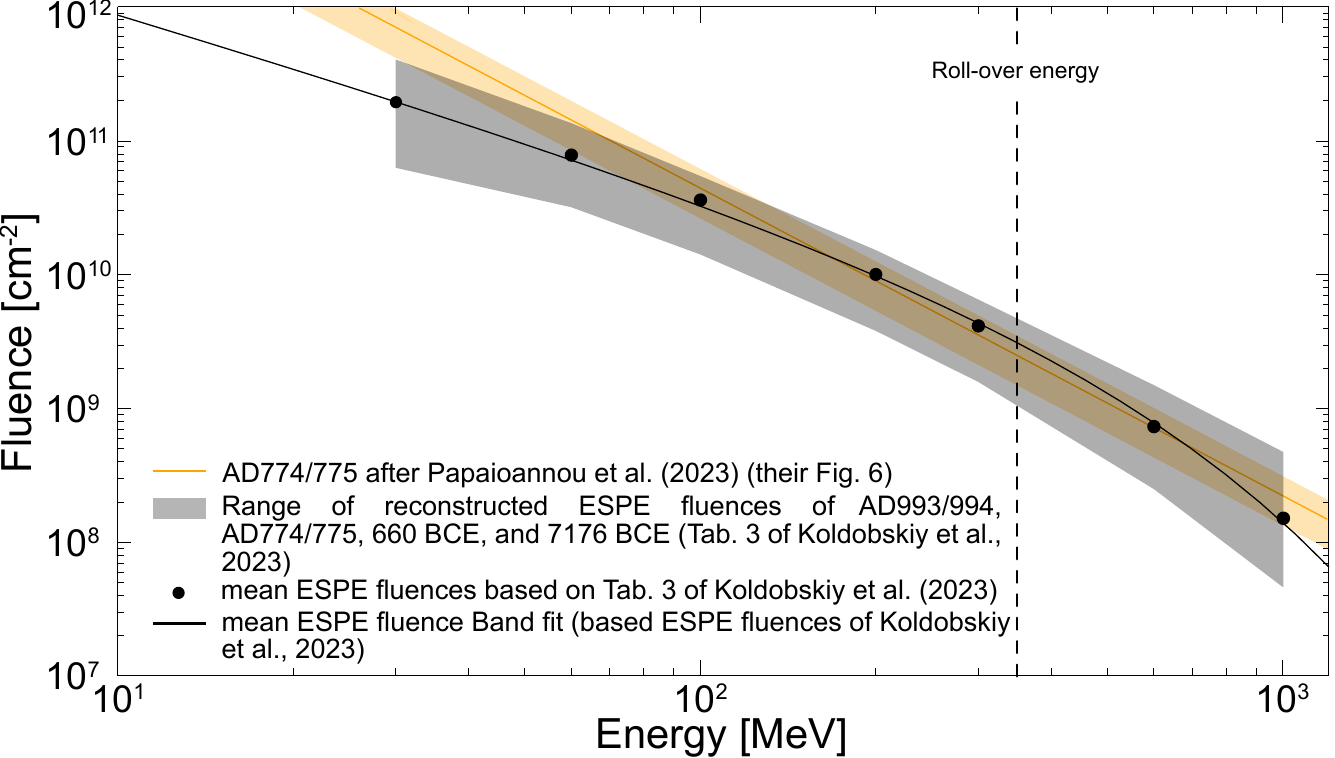}
\caption{The integral fluence spectra of the AD774/775 event taking into account the reconstructed ESPE events around AD993/994, AD774/775, 660 BCE, and 1776 BCE. Here, the gray shade represents the spread of integral fluence values presented in \cite{2023JGRA..12831186K} while the filled circles show the integral energy-dependent mean of each fluence bin, and the black solid line represents the corresponding  Band fit. The orange solid line and envelope are similar to Fig.~\ref{fig:1}.}
\label{fig:2}
\end{figure}
Figure~\ref{fig:2} shows the range of the reconstructed ESPE integral fluence values per integral energy of AD993/994, AD774/775, 660 BCE, and 7176 BCE \citep[gray shaded area, based on][]{2023JGRA..12831186K}. This was constructed using the four values (of the four events) for each integral energy and identifying the range of fluence values per integral energy (represented by the grey area in Fig. \ref{fig:2}). Each vertical column led to a mean value per integral energy. The black dots represent these values. Note that the upper limit is formed by the 660 BCE event (E<15 MeV) and the 7176 BCE event (E>15 MeV), while the AD993/994 event forms the lower limit at all energies. To derive an estimation of the mean of the range of integral fluences per integral energy, we utilize the Band fit function, a double power-law with an energy-dependent roll-over \citep[e.g.,][]{Band-etal-1993}. Here, the omnidirectional event-integrated integral fluences $J(>R)$ [cm$^{-2}$] are given by
\begin{eqnarray}
\small
F(>R) = 
\begin{cases} J_0 \left(\frac{R}{1~\mathrm{GV}}\right)^{-\gamma_1} \exp{\left(-\frac{R}{R_0}\right)} & R \leq (\gamma_2 - \gamma_1)R_0\\ 
    J_0 \left(\frac{R}{1~\mathrm{GV}}\right)^{-\gamma_1} \exp{\left(-\frac{R}{R_0}\right)} \left(\frac{R}{R_1}\right)^{-\gamma_2}& R > (\gamma_2 - \gamma_1)R_0,
\end{cases}
\normalsize
\end{eqnarray}
where $J_0$ represents the fluence normalization coefficient, and $\gamma_1$ and $\gamma_2$ represent the low/high rigidity power-law indices \citep[e.g.,][]{2018JSWSC...8A...4R}. Accordingly, the event-integrated differential spectrum (in units of [1/(cm$^2$ sr GeV)]) can be derived by:
\begin{eqnarray}
\small
    F(>E) = 
	\begin{cases} \frac{1}{4\pi} J_0 \left(\frac{R}{1~\mathrm{GV}}\right)^{-\gamma_1} \exp{\left(-\frac{R}{R_0}\right)} \frac{(\gamma_1R_0+R)(E+E_0)}{R_0 R^2}& R \leq (\gamma_2 - \gamma_1)R_0\\ 
   \frac{1}{4\pi} J_0 \cdot A \cdot \gamma_2 \cdot R^{-\gamma_2} \frac{E+E_0}{R^2}& R > (\gamma_2 - \gamma_1)R_0.
    \end{cases}
\label{eq2}
\normalsize
\end{eqnarray}
Here, $A = \left[(\gamma_2-\gamma_1)R_0\right]^{\gamma_2-\gamma_1}\exp{(\gamma_1 - \gamma_2)}$ and $E_0$ is the protons rest mass energy (0.938 GeV). The following parameters have been found to fit the values (i.e., the black dots in Fig.~\ref{fig:2}) best: $J_0$ = 8.5$\cdot 10^{13}\pm 1 \cdot 10^{11}$ cm$^{-2}$, $\gamma_1$ = 0.65$\pm$0.02, $\gamma_2$ = 7.0$\pm$0.05, and $R_0 = 0.35\pm0.01$ GV. The corresponding roll-over energy is displayed as the black dashed line. 

To test whether the ``worst case" integral fluence spectra inferred from $F_{SXR}$ can further be applied to modern events, we turn to the strongest SEP events that have been recorded at Earth as GLEs above the ever-present background of galactic cosmic rays \citep[see a recent review on the subject in][and references therein]{macau_mods_00083703}. Therefore, the published integral fluence spectra of the GLEs between 1956 and 2017 (i.e., covering GLE55 -- GLE71\footnote{e.g., \url{https://gle.oulu.fi/}}) by \citet{koldobskiy2021new} were used (see Table \ref{tab:1} for more detail). 

Further, we utilized the newly re-calibrated $F_{SXR}$ values for the saturated GOES events \citep{hudson2024} and scaled the rest by utilizing a multiplicative factor of 1/0.7 \citep[see also][]{2023A&A...671A..66P}. Table \ref{tab:1} provides all information on the GLEs employed and gives their initial $F_{SXR}$ and the newly re-calibrated values ($F_{SXR_{rec}}$). Thereby, the lowest $F_{SXR_{rec}}$ value used in our estimations is 7.2$\cdot 10^{-5}$ W/m$^2$ (M7.2) associated to GLE71 on 17 May 2012 \citep{2014SoPh..289.3059R}, while the highest $F_{SXR_{rec}}$ used is 2.57$\cdot 10^{-3}$ W/m$^2$ (X25.7) associated to GLE65. Based on these $F_{SXR_{rec}}$ values, a band of the ``worst case" integral fluence spectra is reconstructed.

\begin{figure}[!t]
\centering
\includegraphics[width=\columnwidth]{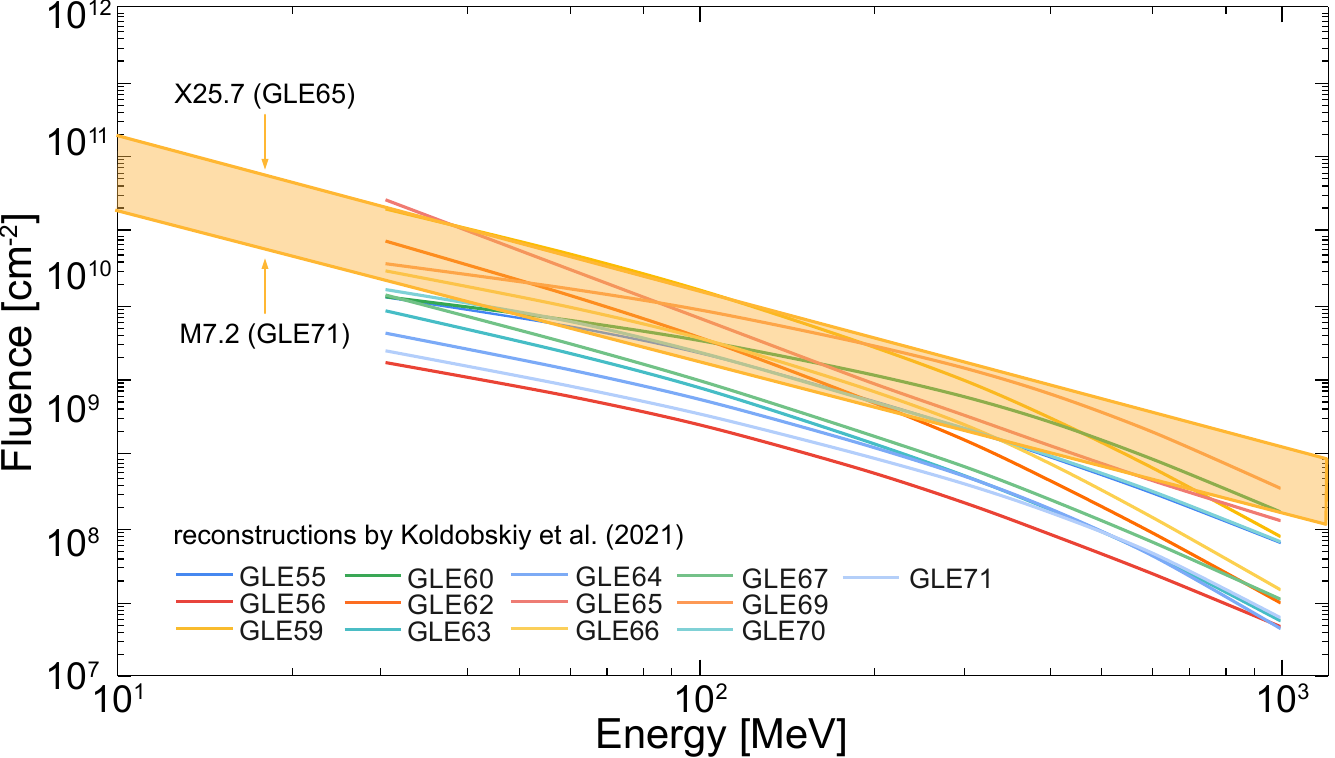}
\caption{The integral fluence spectrum for GLEs from 1956-2017. Each line is a Band fit (see Eq.~(\ref{eq2})) to the values presented in Table \ref{tab:1}. The orange range depicts the ``worst case" estimated integral fluence spectrum for the lower and the highest re-calibrated $F_{SXR}$ values of the associated solar flares for GLE71 (lower) and GLE65 (higher). Note that the results are represented by their obtained PL fit, per case, excluding the according error estimates.}
\label{fig:3}
\end{figure}

The results are shown in Fig. \ref{fig:3} presenting the integral fluence spectra obtained for each GLE \citep{koldobskiy2021new} combined with the range of the ``worst case" integral fluence spectra obtained from the method outlined in \cite{2023A&A...671A..66P} (orange band). Again, our "worst case" integral fluence spectrum agrees well with the integral fluence spectra of the GLEs between 1956 and 2017. This suggests that the method outlined in \cite{2023A&A...671A..66P} can reliably estimate the ``worst case" integral fluence spectra up to $\sim$E>1 GeV - taking into account the limitations of the spectral form discussed here above and in Appendix \ref{appendix:A}.

\section{Conclusions}
The extremes of solar energetic particles were revealed only a decade ago \citep{2012Natur.486..240M} and brought a rapid change in the field since it created many more questions than answers \citep[see the detailed discussion in][]{2023SSRv..219...73U}. Assuming that our host star, the Sun, drives such events and that those represent a high energy/low probability tail of the regular SEP distribution \citep{2021GeoRL..4894848U}, it is desirable to be able to estimate their resulting ``worst case" integral fluence spectra to more reliably estimate the direct technological and societal consequences of such events. 

In this direction, we employed the methodology detailed in \cite{2023A&A...671A..66P} and showed that it is possible to accurately estimate the integral fluence spectra of one of the strongest extreme SEP events ever observed (i.e., AD774/775, Fig.~\ref{fig:1}). The concept was further tested against a mean ESPE obtained based on four extreme SEP events found in cosmogenic radionuclide records \citep[i.e., AD993/994, AD774/775, 660 BCE, and 7176BCE; ][see Fig.~\ref{fig:2}]{2023JGRA..12831186K}. It was additionally tested on a series of strong measured SEP events recorded at the surface of the Earth, i.e., GLEs (Fig.~\ref{fig:3}).  Our derived "worst case" integral influences seem to hold true for all cases, providing a robust upper limit estimation for all (extreme) solar events.  

AD774/775 is the only ESPE event for which a solar association, in terms of SXRs, has been possible \citep{2022LRSP...19....2C}. In addition, the comparison in Fig.~\ref{fig:3} between the mean-ESPE spectrum (black line) and the derived spectrum from AD774/775 (orange line) is suggestive of the magnitude of flares that lead to ESPEs, which should be in the range of a few X100, and that the concept presented here could provide an estimate of the spectrum for ESPEs in case the triggering SXR flux is known.

Our method assumes a direct scaling relation between SXR flux $F_{SXR}$ and the obtained integral fluence $F(>E)$ \citep[see details in][]{2023A&A...671A..66P}. In this work, we utilized for the first time the accurate/scaled $F_{SXR}$ flux \citep[see details in][]{hudson2024}. Our method does not underpin the physical processes involved in the acceleration and propagation of such extreme SEP events. Instead, it directly estimates and quantifies the effects of these changes. The verification of the concept provides valuable insight: even if we do not yet know how these extreme events are unleashed by the Sun and if these are ``black swans" (produced by physical processes already known) or ``red dragons" (produced by processes completely unknown) \citep[see, e.g.,][]{2022LRSP...19....2C, 2023SSRv..219...73U}, we do know that such events will strike again letting our interconnected technologically dependent society completely vulnerable. 

With upcoming missions such as PLATO \citep{RauerEA2014,2024arXiv240605447R}, searching for terrestrial planets around, e.g., G-type and Sun-like stars, it is further essential to constrain the stellar particle and radiation environment to be expected within the habitable zone of such systems \citep[e.g.,][]{EngelbrechtEA2024, HerbstEA2024}.  Caution is needed when a scaling relation is established for a particular system and used for another. Uncertainty arises from differences in the stellar conditions compared to our Sun (e.g., temperature, density, magnetic field) and the stellar/solar wind properties that affect particle acceleration and propagation dynamics. Furthermore, the level of activity of a star versus our Sun, in terms of flare and CME occurrence, directly affects the SXR-SEP relations \citep[see also the discussion in][]{2023A&A...671A..66P}. Assuming the Sun to be an ordinary G-type star, our method also provides a first-order approximation of the "worst-case" integral stellar energetic particle fluence of an active solar twin at the exoplanetary locations and with that paths the way for sophisticated studies on the impact of cosmic rays on exoplanetary atmospheres (i.e., climate and chemistry) and with that on transmission spectra and biosignatures such as methane and ozone.

\begin{acknowledgements}
KH and AP acknowledge the International Space Science Institute and the supported International Team 441: \textit{High EneRgy sOlar partICle Events Analysis (HEROIC)} and Team 464: \textit{The Role Of Solar And Stellar Energetic Particles On (Exo)Planetary Habitability (ETERNAL)}. KH further acknowledges the support of the DFG priority program SPP 1992 “Exploring the Diversity of Extrasolar Planets (HE 8392/1-1)”. AP acknowledges NASA/LWS project NNH19ZDA001N-LWS. This project has received funding from the Research Council of Norway through the Centres of Excellence funding scheme, project number 332523 (PHAB). Part of this work has received funding from the European Union’s Horizon Europe programme under grant agreement No 101135044 (SPEARHEAD). The research was partly funded by the European Union (ERC, PastSolarStorms, grant agreement n° 101142677). Views and opinions expressed are however those of the author(s) only and do not necessarily reflect those of the European Union or the European Research Council. Neither the European Union nor the granting authority can be held responsible for them.
\end{acknowledgements}


\bibliographystyle{aa}
\bibliography{references}


\begin{appendix}
\section{Soft X-ray and CME relations}\label{appendix:A}
Observations and theoretical studies have established a significant relationship between coronal mass ejections (CMEs) and the thermal energy release traced by associated solar flare soft X-ray (SXR) emissions \citep{2020SSRv..216..131P}. Statistical analyses indicate that most CMEs exhibit some level of X-ray emission \citep{2006ApJ...650L.143Y}. However, solar flares occur more frequently than CMEs, suggesting a lack of one-to-one correspondence. Nonetheless, the likelihood of a CME being associated with a flare increases with flare intensity in terms of SXR flux \citep[see Fig. 1 of][]{2006ApJ...650L.143Y}.

A moderate correlation has been identified between CME speeds and the peak SXR flux of associated flares \citep{2015SoPh..290.1337S, 2016ApJ...833L...8T, 2024A&A...690A..60P}. This correlation arises because solar eruptive events — flares and CMEs — are interconnected, both resulting from magnetic field restructuring in the corona. However, significant scatter in the data \citep[see e.g., Fig. 3 in][]{2024A&A...690A..60P} suggests that while SXR emissions relate to CME kinematics, additional factors influence CME dynamics.

Several studies have explored scaling relations between CME properties and SXRs. For instance, \cite{2016ApJ...833L...8T} found that CME speed ($V_{CME}$) scales with the SXR photon flux ($F_{SXR}$) as $V_{CME} \propto F_{SXR}^{0.30\pm0.04}$, implying that more intense flares are generally associated with faster CMEs. Expanding on this, upper-limit scaling relations for CME speeds have been derived based on SXR emissions \citep{2024A&A...690A..60P} and active region (AR) energy \citep[see e.g.,][]{2018eeg..book...37G}. More recently, \cite{2024A&A...691L...8M} applied the concept of scaling laws to investigate analogous correlations, extending the concept of these relations to stellar flares and CMEs.

While these scaling laws provide valuable insights into CME-SXR connections, significant variability remains. Factors such as the surrounding magnetic environment, CME geometry, and projection effects contribute to the observed scatter. Thus, although SXR measurements serve as crucial indicators of CME dynamics, they represent only one aspect of a complex interplay of solar eruptive processes \citep[see the relevant discussion in][]{2024A&A...690A..60P}.


\section{Catalog}\label{appendix:B}
\begin{table*}[hbt!]
\scalebox{0.71}{
    \centering
    \begin{tabular}{c|c|c|c|c|c|c|c|c|c|c|c|c|c}

     E > MeV	& GLE55 & 	GLE56& GLE59&	GLE60&	GLE62&	GLE63&	GLE64&	GLE65&	GLE66&	GLE67&	GLE69&	GLE70&	GLE71\\
      & F [cm$^{-2}$]& F [cm$^{-2}$]& F [cm$^{-2}$]& F [cm$^{-2}$]& F [cm$^{-2}$]& F [cm$^{-2}$]& F [cm$^{-2}$]& F [cm$^{-2}$]& F [cm$^{-2}$]& F [cm$^{-2}$]& F [cm$^{-2}$]& F [cm$^{-2}$]& F [cm$^{-2}$]\\
     \hline
     29	&1.22E+08&	1.61E+07&	1.89E+09&	1.24E+08&	6.98E+08&	8.02E+07&	4.01E+07&	2.51E+09&	2.77E+08&	1.31E+08&	3.46E+08&	1.55E+08&	2.32E+07\\
     60& 4.95E+07& 5.65E+06& 4.68E+08& 5.98E+07& 1.30E+08& 2.15E+07& 1.29E+07& 2.91E+08& 9.00E+07& 2.90E+07& 1.59E+08& 5.79E+07& 7.79E+06\\
    103& 2.06E+07& 2.22E+06& 1.44E+08& 3.08E+07& 3.27E+07& 6.95E+06& 4.89E+06& 5.92E+07& 3.25E+07& 8.66E+06& 7.95E+07& 2.10E+07& 3.12E+06\\
    196& 5.09E+06& 5.48E+05& 2.75E+07& 1.13E+07& 4.92E+06& 1.37E+06& 1.21E+06& 8.87E+06& 6.92E+06& 1.73E+06& 2.80E+07& 4.96E+06& 8.65E+05\\
    294& 1.93E+06& 1.93E+05& 8.50E+06& 5.27E+06& 1.33E+06& 4.45E+05& 4.40E+05& 2.91E+06& 2.15E+06& 5.91E+05& 1.28E+07& 1.96E+06& 3.50E+05\\
    407& 8.24E+05& 7.52E+04& 2.76E+06& 2.46E+06& 3.92E+05& 1.61E+05& 1.65E+05& 1.18E+06& 6.65E+05& 2.20E+05& 5.88E+06& 8.62E+05& 1.48E+05\\
    602& 2.79E+05& 2.30E+04& 6.14E+05& 8.04E+05& 8.17E+04& 4.14E+04& 4.02E+04& 4.14E+05& 1.32E+05& 6.14E+04& 1.87E+06& 2.96E+05& 4.36E+04\\
    1000& 5.96E+04& 4.44E+03& 7.25E+04& 1.56E+05& 9.27E+03& 5.31E+03& 4.18E+03& 1.19E+05& 1.39E+04& 1.04E+04& 3.24E+05& 6.18E+04& 5.85E+03\\
    \hline
    \hline
    F$_{\rm{SXR}}$ [W/m$^2$]& 9.40E-04& 1.10E-04& 5.70E-04& 1.44E-03& 1.00E-04& 7.10E-05& 3.10E-04& 1.70E-03& 1.00E-03& 8.30E-04& 7.10E-04& 3.40E-04& 5.10E-05\\
    F$_{\rm{SXR_{rec}}}$ [W/m$^2$]& 1.34E-03& 1.57E-04& 8.14E-04& 2.11E-03& 1.43E-04& 1.01E-04& 4.43E-04& 2.57E-03& 1.55E-03& 1.33E-03& 1.02E-03& 4.86E-04& 7.29E-05\\\

    \end{tabular}
    }
    \caption{GLEs presented in Fig.\ref{fig:3}. The integral fluence per F(>E) [cm$^{-2}$] for a set of 8 integral energies for 12 GLEs, as published by \cite{koldobskiy2021new}. Associated solar flare flux $F_{XSR}$ [W/m$^2$] followed by $F_{XSR_{rec}}$ [W/m$^2$] which are the currently accurate SXR fluxes. For these values, proper scaling was applied (using a multiplicative factor of 1/0.7 for the GOES 1–8  channel). The saturated strong X-class event values for GLE60, GLE65, GLE66, GLE67, GLE69 and GLE72 also need further attention; their re-scaled SXR classes are taken from \cite{hudson2024}.} 
   \label{tab:1}

\end{table*}


\section{Spectral shape} \label{appendix:C}
Flares and CMEs are widely recognized to almost always coincide during strong SEP events, which typically obscures the primary acceleration mechanisms \citep[see][and references therein]{2022SoPh..297...53K}. Thus, double power-law spectra in SEPs are expected. Such spectra are frequently observed, spanning from the lowest proton energies up to several hundred MeV \citep[e.g.][]{2012SSRv..171...97M}. They are commonly attributed to: (a) the presence of two distinct accelerators—such as CME-driven shocks and flare-related processes. In particular, the assumption is that CME-driven shocks are the dominant acceleration mechanism at lower energies \citep[see, e.g.][]{desai2016large} and flare-associated processes, such as magnetic reconnection, dominate at higher energies \citep{2001SSRv...95..215K}. As a result, below the break energy, the spectrum exhibits a flatter slope (shock acceleration), while above it, the slope becomes steeper (flare-related acceleration); and (b) low-energy particles are more susceptible to scattering and energy losses due to interactions with the solar wind and interplanetary medium, including Coulomb collisions and adiabatic deceleration, which further modify their spectrum \citep{2000ApJ...537.1073D} while high-energy particles, being less influenced by these processes, tend to retain spectral features closer to their source \citep{1999SSRv...90..413R}.  Nonetheless, it is also possible for a single accelerator to produce such spectra under specific conditions: (i) CME-driven shocks can produce a double power-law spectrum if the acceleration efficiency varies across different energy ranges. For instance, the presence of stronger turbulence at higher energies or a limited supply of low-energy seed particles can naturally lead to a change in the spectral slope \citep{1985ApJ...298..400E}; (ii) shock geometry and evolution can also affect particle acceleration. As the shock expands and weakens, it may favor acceleration at certain energy ranges over others, leading to a break in the spectrum \citep{2024A&A...682A.106K}; (iii) even if the initial spectrum from a single accelerator is a pure power law, transport effects such as scattering, energy loss, and escape from the shock region can create a double power-law signature. For example, Coulomb collisions can deplete low-energy particles, while interplanetary propagation can modify high-energy particle distributions \citep{2000ApJ...537.1073D}; (iv) a single mechanism, such as stochastic acceleration (second-order Fermi processes), can inherently produce a spectrum with a break if the energy gain per interaction or the rate of particle injection into the process changes with energy \citep{2012SSRv..173..535P} and (v) flare-accelerated electrons and protons are injected both downward into flare loops and upward into a forming flux rope. These particles remain trapped until the expanding flux rope reconnects with an open coronal structure, releasing them into interplanetary space. Within the dense flux rope, Coulomb collisions reduce the low-energy portion of the proton spectrum. Meanwhile, the CME-driven shock independently accelerates a population of suprathermal particles present in the corona and interplanetary medium, gradually increasing their energy spectrum and replenishing the lower-energy component. Together, these processes result in a double power-law proton spectrum, characterized by a flatter slope below the break energy and a steeper slope above it \citep{2013ApJ...771...82M}. Thus, the spectra of low-energy and high-energy particles in SEPs are typically not aligned, reflecting their origins from possibly different acceleration processes and the subsequent propagation effects.

\end{appendix}
\end{document}